# Longitudinal Prognosis of Parkinson's Outcomes using Causal Connectivity

Cooper J. Mellema[1,2,6]
Kevin P. Nguyen[1,2,6]
Alex Treacher[1,5,6]
Aixa Andrade Hernandez[1,2,6]
Albert A. Montillo, PhD[1,2,3,4,6]
[1]Lyda Hill Department of Bioinformatics
[2]Biomedical Engineering Department
[3]Advanced Imaging Research Center
[4]Radiology Department
[5]Biophysics Department
[5]University of Texas Southwestern Medical Center
Cooper.Mellema@UTSouthwestern.edu
Albert.Montillo@UTSouthwestern.edu

## 1 ABSTRACT

Parkinson's disease (PD) is a movement disorder and the second most common neurodengerative disease but despite its relative abundance, there are no clinically accepted neuroimaging biomarkers to make prognostic predictions or differentiate between the similar atypical neurodegenerative diseases Multiple System Atrophy and Progressive Supranuclear Palsy. Abnormal connectivity in circuits including the motor circuit and basal ganglia have been previously shown as early markers of neurodegeneration. Therefore, we postulate the combination patterns of interregional dysconnectivity across the brain can be used to form a patient-specific predictive model of disease state and progression in PD. These models, which employ connectivity calculated from noninvasively measured functional MRI, differentially predict between PD and the atypical lookalikes, predict progression on a disease-specific scale, and predict cognitive decline. Further, we identify the connections most informative for progression and diagnosis. When predicting the one-year progression in the Movement Disorder Society-sponsored revision of the Unified Parkinson's Disease Rating Scale (MDS-UPDRS) and Montreal Cognitive assessment (MoCA), mean absolute errors of 1.8 and 0.6 basis points in the prediction are achieved respectively. A balanced accuracy of 0.68 is attained when distinguishing idiopathic PD versus the lookalikes and healthy controls. We additionally find network components strongly associated with the prognostic and diagnostic tasks, particularly incorporating connections within deep nuclei, motor regions, and the Thalamus. These predictions, using an MRI modality readily available in most clinical settings, demonstrate the strong potential of fMRI connectivity as a prognostic biomarker in Parkinson's disease.

## 2 INTRODUCTION

Neurodegenerative diseases remain difficult to diagnose and the longer term trajectory remains challenging to prognose, with unbiased assessments being time consuming and difficult to administer. One promising avenue for making informative prognostic and diagnostic decisions for patients is the application of unbiased neuroimaging techniques. Functional Magnetic Resonance Imaging (fMRI) is of particular interest in making diagnostic and prognostic decisions. Functional changes are an early and

A summary of all acronyms used in this paper is presented in **Supplemental Section 10.1.**

sensitive marker of neurodegeneration, preceding atrophy visible in structural MRI and diffusion MRI (Dadi et al., 2019), not requiring expensive and ionizing radiation like PET and SPECT imaging (K. H. Leung et al., 2019; Mansu et al., 2017), and providing insights into the pathophysiology of these complex, multifacted diseases (Noble et al., 2019; Smitha et al., 2017). Derived measures from fMRI, particularly interregional brain connectivity, has been shown to be strongly predictive in a wide variety of prognostic and diagnostic analyses for neurodegenerative diseases as well as providing insights into the networks and signaling pathways associated with disease symptoms, prognoses, and subtypes (Smitha et al., 2017). The second most common neurodegenerative disease is Parkinson's disease (PD) (Jankovic and Tan, 2020; Pringsheim et al., 2014), a movement disorder with no clinically accepted neuroimaging measures of prognosis. PD also has no clinically accepted tool to distinguish between PD and the similar neurodegenerative diseases: Multiple System Atrophy (MSA) and Progressive Supranuclear Palsy (PSP). In order to fill this unmet need, we develop predictive models using newly proposed measures of connectivity to predict an individual's long term prognosis including the decline in both the motor and cognitive domains – achieving state of the art performance in both of these predictive domains. We also build a model to differentiate PD from atypicals and controls. We also extend a data augmentation technique to fMRI for more accurate machine learning predictors. Furthermore, we use a dataset with minimal medication confounds to not only make more accurate predictions, but also analyse both predictive functional connectivity features and effective connectivity features associated with progression. The tools and analyses presented in this manuscript are a promising step towards addressing these clinical needss and the implicated connections provide new avenues for further research and potential therapeutic targets.

### 2.1 PD diagnosis

Parkinson's disease is the second most common neurodegenerative disease with poorly understood functional changes as the disease progresses (Jankovic and Tan, 2020; Pringsheim et al., 2014). PD has two major symptomatic lookalikes, PSP and MSA. Despite the similar symptoms, these diseases have a distinct pathophysiology and require distinct treatment (Andrew J. Hughes, Susan E. Daniel, Yoav Ben-Shlomo, Andrew J. Lees, 2002). The current clinical diagnostic accuracy is approximately 87% in movement disorder specialist clinics, but is typically only 63% in general settings (Andrew J. Hughes, Susan E. Daniel, Yoav Ben-Shlomo, Andrew J. Lees, 2002). As 63% balanced accuracy for general clinics is rather low, there has been great interest in developing machine-learning based diagnostic models to supplement clinical findings and aid diagnosis. Past studies have achieved diagnostic balanced accuracies with machine learning in the multitask prediction diagnosing between normal control (NC) versus PD versus PSP versus MSA of up to 0.77 (Chougar et al., 2021). We contribute to the diagnostic problem by mining our predictive models for connectivity correlates in order to better elucidate the pathophysiology of PD.

### 2.2 PD prognosis

Predicting the prognosis of PD subject would inform patients and allow researchers to inform clinical trials with expected progression, accelerating efforts to find effective treatments. Currently, there are some heuristics that currently guide PD prognosis including the unilaterality versus bilaterality of symptoms and age of onset, but these remain coarse and non-quantitative (Jankovic and Tan, 2020). Identification of biomarkers to predict progression are needed (Marek et al., 2018). The minimal clinically important difference (CID) in PD patient as measured by UPDRS is 2.7, a moderate CID is 6.7, and a large CID is 10.7 (Lisa M. Shulman, Ann L. Gruber-Baldni, Karen E. Anderson, 2010). Todays methods achieve MAEs as low as 3.22, missing the ideal minimal clinically important difference of 2.7. Previous studies which employ neuroimaging biomarkers have successfully employed T1 imaging (Zeighami et al., 2019), SPECT imaging (K. H. Leung et al., 2019; Mansu et al., 2017), and DTI imaging (Taylor et al., 2018) to predict PD progression. They do not meet the minimal CID threshold and do not offer additional connectivity insights into the pathophysiology of the disease. However, there has also been evidence of early and sensitive prognostic signal or correlates in fMRI. Unfortunately, the majority of these studies are confounded by the patients being on medication at the time of recording and MDS-UPDRS evaluation. Previously identified functional correlates to diagnosis, specific symptom severity, medication response, and other non-prognostic neuroimaging biomarkers are numerous in the literature, suggesting a robust functional signal indicating PD pathophysiology (Amboni et al., 2015; Baggio et al., 2015; Hassan et al., 2017; Hou et al., 2017; Ng et al., 2017; Tuovinen et al., 2018). Specific functional correlates of long-term prognosis as measured with MDS-UPDRS or MoCA have been found by numerous researchers, confirming that there are not only correlates to current severity, but also future severity (Burciu et al., 2016; Hou et al., 2017; Kim T.E. Olde Dubbelink, Menno M. Schoonheim, Jan Berend Deijen, Jos W.R. Twisk, Frederik Barkhof, Henk W. Berendse, 2014; Manza et al., 2016; Simioni et al., 2016). However, there has been a dearth in using effective connectivity (i.e. causal connectivity) as a prognostic predictor. Our study quantifies the predictive power of effective connectivity for prognosis in PD. Effective connectivity can have higher predictive power than functional connectivity, but more importantly effective connectivity has a more direct interpretability in the discovered connections and allows us greater insight into pathophysiology than purely functional measures (Bielczyk et al., 2019; Chockanathan et al., 2019; Friston et al., 2019; Mellema and Montillo, 2022; Yao et al., 2017).

**Table 1: PDBP Demographics**. Part **1a** shows the demographics of subjects with an available fMRI and diagnosis, Part **1b** shows the demographics of subjects with an available initial fMRI and recorded 1 year change in MDS-UPDRS score, and part **1c** shows the demographics of subjects with an available initial fMRI and recorded 1 year change in MoCA.

| | ***Target*** | ***# subjects*** | ***Age*** | ***% Female*** | ***# PD/NC/PSP/MSA*** | |
|---|---|---|---|---|---|---|
| ***a*** | Diagnosis | 146 | 65.1±8.8 | 42.5% | 73/45/21/7 | |
| | | | | | **Initial score** | **Δ score** |
| ***b*** | MDS-UPDRS | 63 | 64.4±9.0 | 31.6% | 40.0±27.4 | 12.5±6.5 |
| ***c*** | MoCA | 71 | 64.0±8.9 | 31.0% | 25.89±2.76 | 0.25±2.24 |

## 2.3 Contributions

This paper contributes several important predictive models and theoretical advancements: (1) First, we develop diagnostic models for PD, achieving excellent balanced accuracy to differentiate PD, PSP, MSA and normal controls. (2) Second we develop state-of-the art prognostic models that predict the 1 year change in motor symptomatology and we develop prognostic models predicting the 1 year change in cognitive decline. (3) Third, we quantitatively compare four measures of functional and effective connectivity, including newly proposed measures which incorporate structural priors as well as traditional measures across the aforementioned diagnostic and prognostic predictive tasks. (4) We develop an anatomically realistic augmentation method for 4D fMRI–derived connectivity analysis and demonstrate how it is a driving factor to significantly improve predictive performance. (5) We identify the most strongly associated connections to the predictive targets, revealing what the models have learned.

## 3 MATERIALS AND METHODS

### 3.1 Materials

This work uses fMRI data from the Parkinson's disease biomarkers program (PDBP) (Ofori et al., 2016) to make diagnostic and prognostic models. The PDBP dataset was chosen as the primary dataset for analysis due to its emphasis on fMRI data collection while patients were off medication, a significant confound in PD outcome studies. Task-based fMRI was gathered by PDBP, but was treated as resting data in subsequent analysis, so that we can apply our connectivity analysis. Previous studies have found extremely high concordance between task and rest fMRI signal (greater than 80% concordance) and have successfully used task fMRI for resting analysis (Beheshtian et al., 2021; Kraus et al., 2021). Connectivity measures were derived from baseline (year 0) fMRI. These measures were used as inputs to our predictive models.

For the diagnostic task, we distinguish PD from other neurodegenerative lookalikes and healthy controls. We predict a four-way differential diagnosis between patients with PD, PSP, MSA, or Normal Control (NC). For the prognostic measures, the change in either a composite measure emphasizing motor symptoms and activities of daily living (MDS-UPDRS) or cognitive score (MoCA) over a period of 1 year was used as the prediction target.

The PDBP dataset includes 73 PD subjects with fMRI. An additional 73 subjects, including subjects with atypical Parkinsonian syndromes and control subjects, were collected. We performed 3 predictive tasks on PDBP: (1) differential diagnosis, (2) regression of the progression in a disease-severity score, and (3) regression of the progression in a cognitive score. Demographics for the subjects used in the subsequent diagnostic prediction experiments is shown in **Table 1a**. For the motor progression task, we use 57 PD subjects who had a recorded one-year change in the MDS-UPDRS after removal of outliers. We removed 6 subjects with a highly negative one-year change in MDS-UPDRS (improvement greater than 10 basis points) as outliers and potentially non-physiologic. PD is generally considered to be a progressive and irreversible disease, so a large improvement greater than typical measurement noise is likely an error in measurement or diagnosis (Jankovic and Tan, 2020; Pringsheim et al., 2014). Demographics for the longitudinal motor progression prediction are shown in **Table 1b**. For the longitudinal cognitive prediction task, we use 71 PD subjects who had a recorded 1 year change in their Montreal Cognitive Assessment (MoCA). Demographics for the longitudinal cognitive progression prediction are shown in **Table 1c**.

### 3.2 Methods

#### 3.2.1 *Data processing*

Clinicodemographic correlates were included in all analysis. These included the site of data collection, age, gender, ethnicity, education level, and handedness.

All fMRI data was prepared with an in-house fMRI preprocessing pipeline with advanced motion correction, shown to be significantly better at removing motion artifact in PD than competitors (Raval et al., 2022). Pipeline processing details are provided in **Supplemental Section 10.2**. Mean regional timeseries were extracted with the Schaefer atlas with 100 cortical regions and 35 additional subcortical regions included (Schaefer et al., 2018). The Schaefer atlas is a functional atlas whose regions are defined through the clustering of functional activity in fMRI. This functional atlas was chosen because: 1) a functional atlas tends to capture better functional variability than a purely anatomical atlas (Mellema et al., 2022), and 2) the Schaefer atlas groups regions into resting-state networks (RSNs), which facilitates inter and intra RSN partitioning and analysis. The cerebellum and striatum from the AAL atlas (Rolls et al., 2020b) were included as well because both structures contain signals of diagnostic importance and are often overlooked in prior analyses (Stoodley et al., 2012).

From each mean regional timeseries, the timeseries was linearly detrended and z-scored normalized, after which a set of functional connectivity (FC) and effective connectivity (EC) measures were derived. The FC measures included Correlation ($Corr$), Partial Correlation ($PCorr$), and a machine learning measure of functional connectivity from an XGBoost predictive model ($ML.FC_{XGB}$). Correlation and partial correlation are commonly used measures of FC, while $ML.FC_{XGB}$ has been shown to potentially contain more predictive information for prognostic tasks than the other two FC measures (Mellema and Montillo, 2022). $ML.FC_{XGB}$ was shown to have the highest reproducibility across multiple scans of the same subject and higher predictive power across physiological and cognitive prediction targets. The EC measures included Principal Components Granger Causality ($PC.GC$), which calculates a Granger-Causality score from a lower-dimensional Principal Components projection, and Structurally Projected Granger Causality ($SP.GC$). $SP.GC$ was proposed in (Mellema and Montillo, 2022) and calculates a Granger-Causality score from a lower-dimensional Structural Projection which uses a diffusion-MRI derived tractography connectivity prior in the initial projection. $SP.GC$ regularizes the brain connectivity by projecting brain activity into a lower dimensional space with a preference for communication along physically connected paths. $SP.GC$ was shown in (Mellema and Montillo, 2022) to have higher reproducibility than other EC measures and having more predictive power for multiple targets. $PC.GC$ also performed well in predictive tasks (Mellema and Montillo, 2022). See **Supplemental Section 10.3** and Mellema et al (Mellema and Montillo, 2022) for further detail.

### **3.2.2** *Augmentation*

Augmentation can improve model performance without introducing spurious findings (Phillip Chlap, Hang Min, Nym Vandenberg, Jason Dowling, Lois Holloway, Annette Haworth, 2021). Augmentation introduces realistic perturbations which cover more of the population variation than exists in the original dataset. We use the BLENDS method which generates realistic perturbations in fMRI by warping to new subject anatomy (Nguyen et al., 2021a). BLENDS assumes that anatomical brain configuration (the gyri and sulci) is not a primary determinant in the disease process. Therefore new fMRI can be constructed from existing by warping existing brain shape to that of another subjects, with the same diagnosis. In PD, atrophy of deep nuclei but not cortical brain shape is associated with early diagnosis or progression (Zeighami et al., 2019), so the BLENDS methodology is appropriate as an augmentation approach.

Currently BLENDS is the one of the few augmentation strategy suitable for deriving connectivity measures as it allows the generation of new timeseriese from the fRMI. However, the magnitude of the BLENDS induced variation is small, leading to an $R^2$ across timeseries of the PDBP diagnostic subject set of 0.79+/-0.02. Therefore, we adapt the BLENDS method by magnifying the perturbation by a gain factor to generate a larger perturbation. In particular, the difference between the original timeseries and initial BLENDS-perturbed timeseries was calculated and the difference was scaled to a gain factor of 0.5, i.e. while the original fMRI timeseries was scaled to 0 mean and unit variance, the difference was scaled to 0 mean and 0.5 variance. This scaled difference was added back to the initial fMRI timeseries to create the augmented timeseries. The augmented timeseries then had connectivity measures calculated as was done in the initial timeseries. The augmentation is illustrated in **Figure 1 and the desirable decrease in connectivity measure impact on the timeseries** $R^2$ down to as low as $R^2$ 0.26 is presented in **Table 2.**

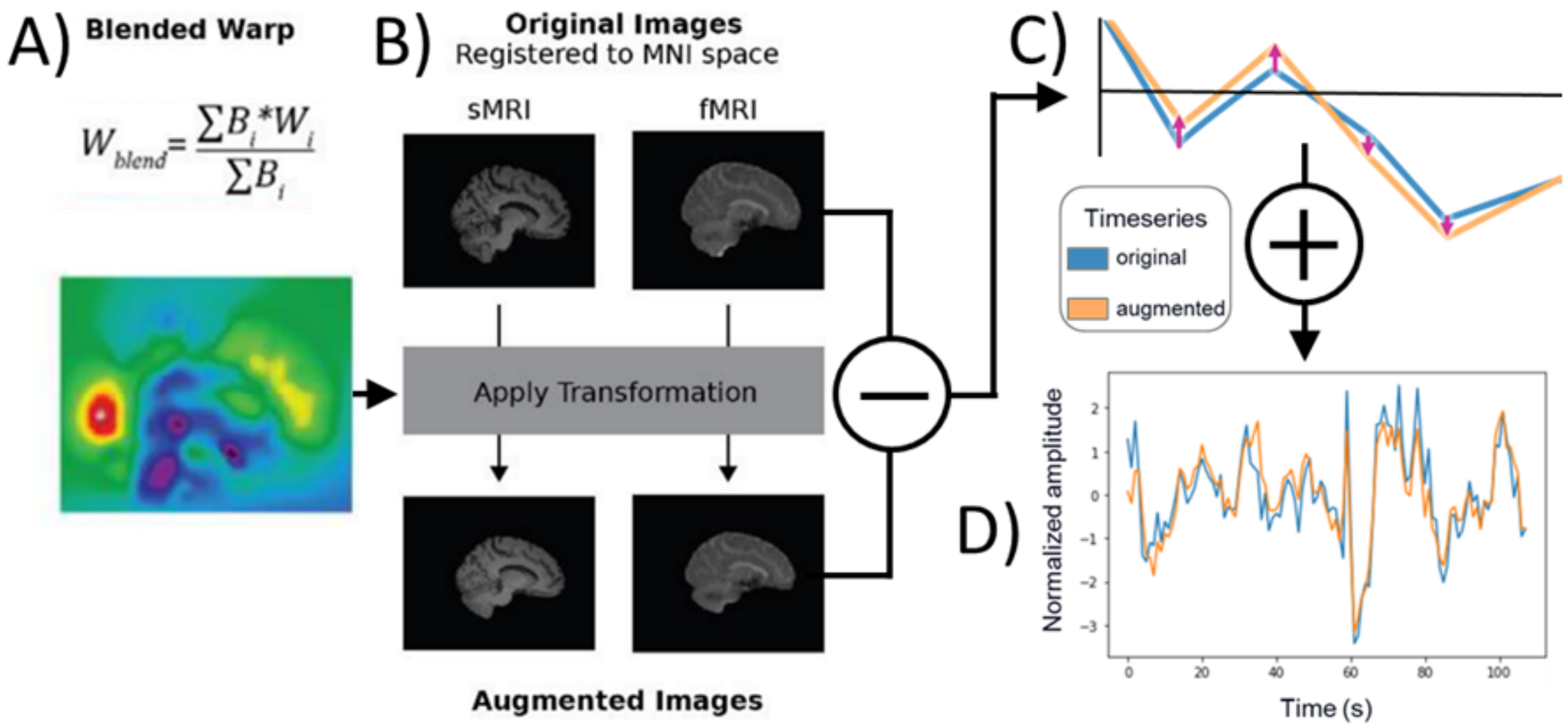

**Figure 1: Timeseries augmentation schematic** shows the overall approach for how the augmentation was performed. First (not pictured) a set of subjects' T1 MRI were nonlinearly warped to MNI-152 space. The composition of the warp to and from MNI space was put into a library of warps, and the warps were blended with a random gaussian mixing of 5 different subjects' warps **(A)**. Next, the original fMRI had this new, blended warp applied to every frame to generate realistic distortions as if the original fMRI activity had been projected onto a combination of different subjects' brans and back to MNI space **(B)**, generating small but realistic distortions in the extracted mean regional timeseries **(C)**. The small differences in the timeseries were subtracted out and added back in with a gain factor to generate larger timeseries differences, treating the first realistic distortion from the cross-anatomical registration as analogous to the initial step in a gradient descent **(D)**.

*To maximize model learning,* augmentation was performed to create a uniform distribution of scores and diagnoses. The original distribution of original training subjects shows a unimodal Gaussian-like distribution of the change in MDS-UPDRS-III and change in MoCA as shown in **Figure S2a** in orange, and the matched and augmented distribution post-augmentation shown in **Figure S2a** in blue. This scheme generates more cases that are farther appart in the motor and cognitive targets and provides more of the 'easier to predict' targets which is related to curriculum learning (Han et al., 2018). $ML.FC_{XGB}$ was not calculated on the additional augmented subjects due to the large increase in compute time and minimal befefit found from using different connectivity measures. Note that to avoid introducing bias in the evaluation metrics, augmentation is performed only on training data and not used for the held out test subjects in our dross validation driven model development.

**Table 2: Similarity post-augmentation** shows the similarities of the measures obtained before augmentation and after augmentation. The $R^2$ for the entire timeseries is on the left, and the $R^2$ for the augmented connectivity measures are on the right.

| *$R^2$ between original and augmented timeseries* | *Connectivity measure* | *$R^2$ between original and augmented connectivity measure* |
|---|---|---|
| *0.79±0.02* | *Corelation* | 0.74±0.05 |
| | *Partial Correlation* | 0.78±0.04 |
| | *PCA Granger Causality* | 0.49±0.23 |
| | *Structurally Projected Granger Causality* | 0.26±0.22 |

#### 3.2.3 *Predictive model training*

We use an XGBoost predictive model for the PDBP predictions (Chen and Guestrin, 2016) as it has been shown to be highly efficacious for tabular data and has shown to be apt to identify predictive signals in other neurodegenerative diseases (Marinescu et al; Torlay et al., 2017). We used a multiple-hold out validation scheme to hyperparameter-optimize and evaluate our model. This training setup has been shown to be more representative of real-world model performance after hyperparameter optimization than a single held-out test set. For an unbiased estimate of model performance we emply a 10x9 nested cross-validation partitioning. The folds were split with stratification by the target (diagnosis, MDS-UPDRS-III, etc.), age, gender, and ethnicity. Within each inner loop, Bayesian hyperparameter optimization (Liaw et al., 2018; Ruben Martinez-Cantin, 2014) was performed with 128 model configurations. Further model optimization details are in **Supplemental Section 10.4**.

For all models, each continuous feature was z-score normalized based on the training foldsdata, while each categorical feature was encoded as an integer. Clinicodemographic covariates including site of data collection, age, gender, ethnicity, education level, and handedness were included as predictors. Missing values of education level and handedness were filled with NaN, while age, ethnicity, and gender were available for all subjects. In addition, FC or EC features were also included as predictors. The clinicodemographic data was preserved for each augmentation of a given subject. As the dimensionality of the Flattened EC or FC matrix is 18,225 for much fewer samples, we dimensionality reduce the large EC/FC matrix with one of two methods: a principal components dimensionality reduction on the EC/FC matrix where we keep a number of components chosen by the 'knee' in preserved variance per componentusing the kneedle algorithm (Satopaa et al., 2011). We also use a univariate dimensionality reduction as an alternative dimensionality reduction technique, preserving the top 5% of the most target correlated edges in the training data.

#### 3.2.4 *Experiments*

*Experiment 1: PD diagnosis*

Our first experiment classifies between patients with PD, PSP, MSA, or Normal Control (NC). We use data from 45 NC patients, 73 PD patients, 21 PSP patients, and 7 MSA patients. We test our ability to differentiate these diagnoses when using clinicodemographic features as well as one of the following EC or FC measures: $Corr$, $PCorr$, $ML.FC_{XGB}$, $PC.GC$, or $SP.GC$. We also test augmentation factors of 0 (no augmentation), matching diagnosis distributions, 5x, 10x, and 20x. We use the XGBoost HPO training scheme as described above, choosing the best model per inner validation fold with the highest balanced accuracy on those inner folds and evaluating its performance in the test partition from the outer fold not seen during model training nor selection. The test performances (balanced accuracy) on each of the 10 outer folds of these best models are reported.

*Experiment 2: PD motor symptom (MDS-UPDRS-III) progression*

Our second experiment regresses the 1 year change in the MDS-UPDRS-III scores in PD patients. This set of MDS-UPDRS scores was acquired while patients were off PD medications for at least 12h. We perform this regression with the aforementioned clinicodemographic features and EC or FC measures, and include baseline MDS-UPDRS-III as a predictor. We test augmentation levels of 0x (no augmentation), 'Match' achieving a uniform distribution over diagnosis, 5x, 10x, and 20x. We train XGBoost models with Bayesian hyperparamter optimization (Liaw et al., 2018; Ruben Martinez-Cantin, 2014) and choose the hyperparameter configuration with lowest mean absolute error on the validation set. We report the test performance on the outer 10 folds of the best performing models on the inner validation folds.

*Experiment 3: PD cognitive symptom (MoCA) progression.*

Our third experiment regresses the 1 year change in the MoCA assessment of 71 PD patients. We use the initial MoCA score as an additional covariate and do not use the initial MDS-UPDRS score, but otherwise follow the same experimental setup as the MDS-UPDRS 1 year regression.

# **4** RESULTS

## 4.1 Experiment 1: PD diagnosis

**Figure 2a** shows the test performance as measured by the balanced accuracy providing a differential diagnosis between PD, PSP, MSA, and PD. Chance accuracy is 0.25. Using the partial correlation connectivity metric at an augmentation factor of 10, when combined with clinicodemographic correlates, achieves the highest balanced accuracy of all combinations: 0.68 using univariate dimensionality reduction and 0.65 with the PCA dimensionality reduction. It also achieved an F1 score of 0.67. We performed a t-test testing the null hypothesis that the distribution of balanced accuracy across all folds is different than the chance accuracy and false discovery rate corrected the calculated p-value with the Benjamini Hochberg procedure at a false-discovery rate of 0.05. The calculated p-value was 7.6E-5 for the PCA case and 7.6E-5 for the univariate case. In general univariate dimensionality reduction provided better performance than PCA dimensionality reduction. Additional performance metrics are presented in **Supplemental Section 10.5**.

## 4.2 Experiment 2: PD motor symptom (MDS-UPDRS-III) progression

The held-out test predictive performance (as measured by MAE) of prognostic XGBoost models predicting the 1 year change in MDS-UPDRS-III is shown in **Figure 2b**. Chance MAE is 3.9 and was calculated by predicting the median change in MDS-UPDRS-III for all subjects. The best (lowest) MAE of 1.8 was achieved by the partial correlation connectivity metric using an augmentation factor of 20 and univariate dimensionality reduction, which also achieved an $R^2$ of 0.69. The corresponding FDR corrected p-value for that best-performer is 6.0E-7 (again testing the null hypothesis that the achieved accuracy is no greater than chance as in Experiment 1). The PCA best-predictor was not as high performing, but achieved an MAE of 2.0, an $R^2$ of 0.59, and a p-value of 2.0E-6. On average across multiple connectivity measures and augmentation factors, the PCA dimensionality reduction technique outperformed the univariate dimensionality reduction technique. **Supplemental Section 10.6** presents additional performance metrics for every predictor.

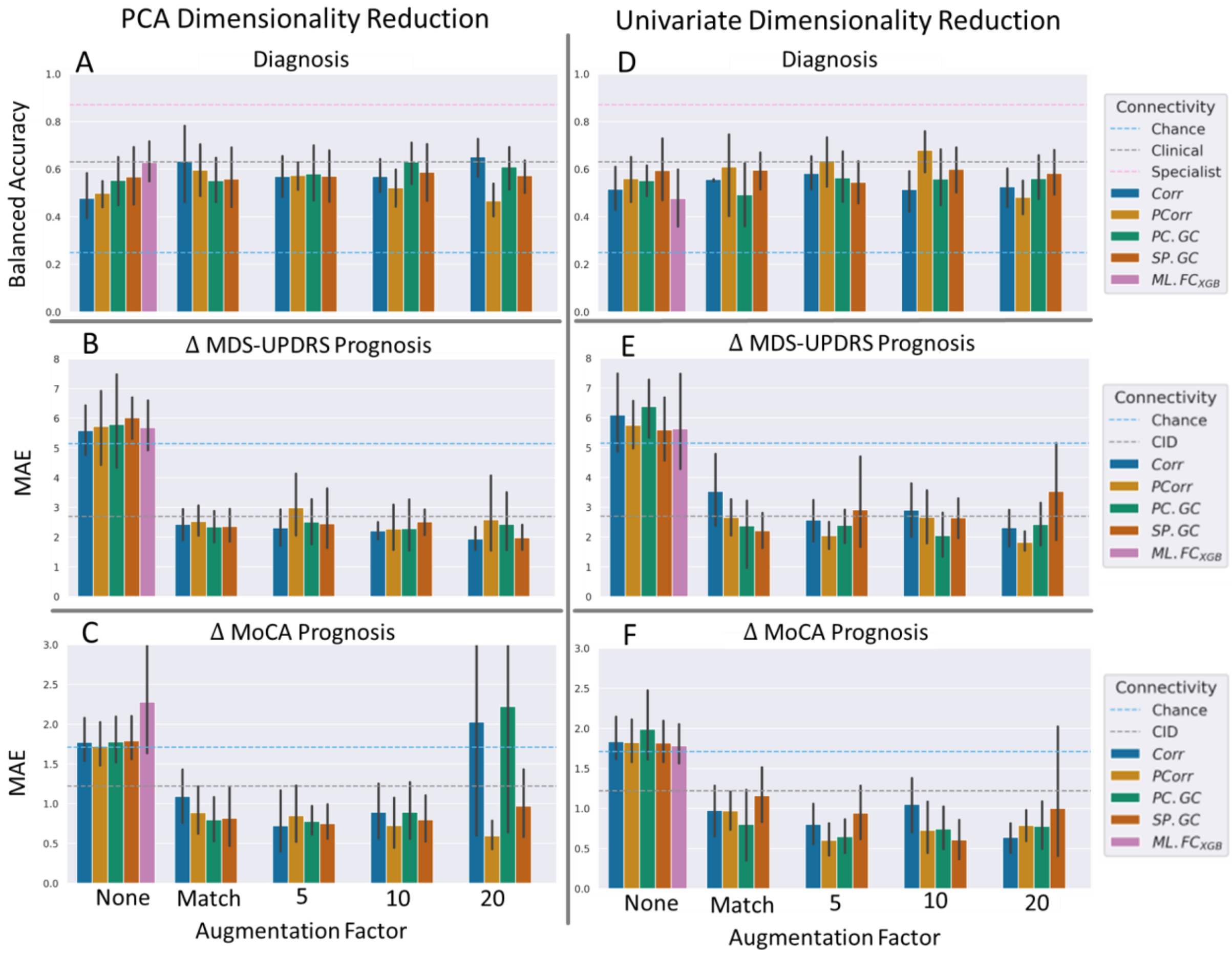

**Figure 2: Diagnosis and prognosis of PD.** This figure shows the performance of models trained to predict various targets with different connectivity features and levels of augmentation. The left column shows the predictive performance of models trained with a PCA dimensionality reduction step while the right column shows the predictive performance of models trained with a univariate dimensionality reduction step. The first row shows the balanced accuracy when making a NC/PD/PSP/MSA diagnostic classification, while the second and third rows show the mean absolute error (MAE) predicting the 1 year change in either MDS-UPDRS or MoCA, respectively. The different colors of bar correspond to the different connectivity feature used as the predictor. The dashed lines show chance accuracy. For the diagnosis tasks, standard and specialist diagnostic accuracies are also shown. For the prognostic task, the clinically important difference (CID) is also shown. **A** shows the balanced accuracy predicting NC/PD/PSP/MSA when using a PCA dimensionality reduction step in the regression. **B** shows the MAE predicting the 1 year change in MDS-UPDRS when using a PCA dimensionality reduction step in the regression. **C** shows the MAE predicting the 1 year change in MoCA when using a PCA dimensionality reduction step. **D** shows the balanced accuracy in the diagnostic prediction when using a univariate dimensionality reduction. **E** shows the MAE for a univariate reduced model predicting the change in MDS-UPDRS. **F** shows the MAE for a univariate reduced model predicting the change in MoCA.

### 4.3 Experiment 3: PD cognitive symptom (MoCA) progression

**Figure 2c** shows predictive performance of models predicting the 1 year change in MoCA on the held-out test sets in the outer multiple-hold out loop. Chance performance was an MAE of 1.6. The lowest MAE of 0.60 across predictive feature sets and an $R^2$ of 0.81 was achieved using a model with the partial correlation connectivity metric, PCA dimensionality reduction and an augmentation factor of 20x. The FDR corrected p-value for that best performer is 4.0E-5. The best EC performance was achieved with the SP.GC features with univariate dimensionality reduction and an augmentation factor of 10x. This approach achieved an MAE of 0.61, R2 of 0.79, and FDR corrected p-value of 8.6E-5. The results of models using the PCA dimensionality reduction overall performed equally as well as the univariate dimensionality reduction. Additional metrics are shown in **Supplemental Section 10.5**.

### 4.4 Augmentation sensitivity analysis

After finding significant benefit to including augmentation in the training of models in Experiments 1-3, we additionally embarked on further analysis of the augmentation benefits with sensitivity studies. Further description and results of these experiments are described in **Supplemental Section 10.6** This supplementary study indicated several findings: We also found the largest benefit from augmentation to be between the factors of 1-3, or 2-4x the original dataset size. This potentially indicates a more ideal number of samples for future diagnostic and prognostic PD studies could be around 150-200 subjects, although an ideal number of 200 augmented samples is not directly equivalent to 200 true subjects.

### 4.5 Feature analysis

To identify the multivariate features most associated with a given outcome, we analyze the important principal components for each of the prognostic tasks, as the PCA reduced models performed better on average than the univariate-reduced models. This allows us to obtain insights into the multivariate associations of both causal and correlative features to the prognostic outcomes. Unlike the prognostic models, the diagnostic model converged on using clinicodemographic features alone in its diagnostic decisionmaking and is therefore not shown.

To maximize the reproducibility of imaging biomarkers, we identify features with high importance across multiple models (Botvinik-Nezer et al., 2020; Mellema and Montillo, 2022). We take the 10 most predictive hyperparameter configurations on each inner loop of training and retrain XGBoost models with these parameters on the entire dataset using the same principal components refitted on the entire dataset. This allows us to estimate distributions of importance across these features. We use the Gini feature importance weighted by the number of samples routed through any given decision node as our feature importance metric for these XGBoost models. We choose to analyze the best FC and best EC connectivity features in greater depth, Partial Correlation and SP.GC. We perform this analysis across both progression tasks ( change in MoCA and change in MDS-UPDRS-III). **Figure 3** show the calculated feature importances across these different tasks, with the clinicodemographic features presented left to right in descending order of importance followed by the connectivity features presented left to right in order of descending importance. Both the Partial Correlation and SP.GC features for both the MoCA and MDS-UPDRS-III tasks use just the first component of the PCA decomposition, and that component is more important to the diagnostic model than any singular featue (not pictured: the univariate association as measured with $R^2$ between the first component and the target varies from 0.02-0.11). **Figure 3A-D** shows that the total importance of the clinicodemographic features exceeds that of the connectivity features, but the connectivity importance is highest in isolation.

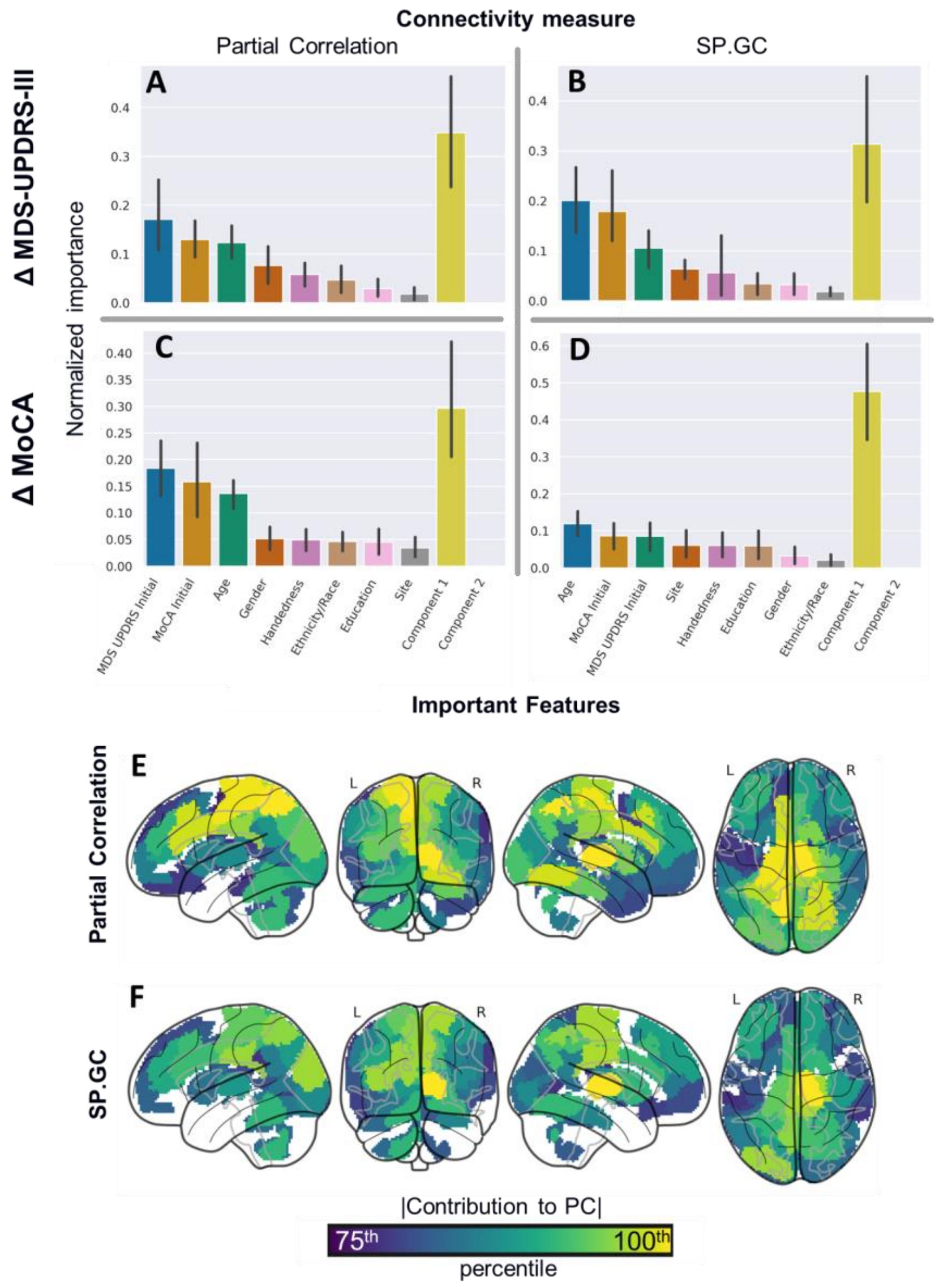

**Figure 3: Feature importances.** This figure shows the importance of features and visualizes the connectivity features used by the prognostic models enumerated earlier. **A-D** show the feature importances (with the 95% CI) of the 10 winning PCA dimensionality reduced models on either the MoCA or MDS-UPDRS-III prediction task from the 10x9 nested cross validation retrained on the entire dataset using either Partial Correlation or SP.GC features, which had the highest average feature importance of the connectivity measures. The PC components were fit across all data, using the same components for the MoCA and MDS-UPDRS-III prediction. **A** shows the feature importances of the models using Partial Correlation features to predict MDS-UPDRS-III. **B** shows the feature importances of the models using SP.GC features to predict MDS-UPDRS-III. **C** shows the features importances of the models using Partial Correlation features to predict MoCA. **D** shows the feature importances of the models using SP.GC features to predict MoCA. **E** and **F** show the weighted degree of each node (ROI) in the connectivity graph of edge contribution to the principal components. **E** shows the weighted degree of the Partial Correlation PC contributions. **F** shows the weighted degree of the SP.GC PC contributions.

As all models converged on using just the first principal component to achieve maximal validation performance, **Figure 3E-F** shows a visualization of the first principal component. For context, each principal component weights every edge in the set of connectivity matrices with a contribution to that component. This visualization condenses that weight per edge of the graph of connections between every ROI to a singular image. This is done by summing the weighted contribution of every edge feeding into and out of a ROI to get a scalar value per region. Equivalently, this is the weighted degree per ROI of the graph representation of the principal component contribution weights. For the Partial Correlation component, we can see edges involving the deep nuclei and thalamus, as well as primary and secondary motor areas are of significant importance in each component. We also see lesser contributions from the occipital lobe and prefrontal cortex. The important regions in this component are similar to the default node network, with the addition of the important deep brain motor nuclei. The SP.GC component has more involvement of edges involving the deep nuclei and thalamus, with lesser contributions from similar regions as the Partial Correlation principal components. The primary difference between Partial Correlation and SP.GC components is the heavier emphasis in the SP.GC component of the deep nuclei. A secondary visualization presenting the top raw edges are shown in **Supplemental Figure S4**. The singular edges suggest that connections from deep nuclei to prefrontal areas are significant in these important principal components. As these full components are associated with outcome targets rather than individual edges, additional work needs to be done to determine which edges from these multivariate combinations of edges are singularly associated with the prognostic targets.

# **5** DISCUSSION

## 5.1 Significance

Overall, we achieve extremely low test MAEs for the prognostic prediction of overall PD severity, as measured by MDS-UPDRS, and cognitive decline, as measured by MoCA. Shulman identified the minimal clinically important difference (CID) in PD patient UPDRS of 2.7, a moderate CID of 6.7, and a large CID of 10.7 (Lisa M. Shulman, Ann L. Gruber-Baldni, Karen E. Anderson, 2010). We have surpassed the minimal CID identified (though longitudinally and with the updated MDS-UPDRS score) and far surpass the moderate and large CIDs. Similarly, Wu identified a minimal CID in MoCA of 1.22 for stroke patients (Wu et al., 2019). Although we are examining a different disease, a similar minimal CID for PD patients has not been identified so we will use this threshold as a reasonable proxy goal. We also meet this minimal CID. We additionally achieve reasonable balanced accuracy across the differential diagnosis of PD and its lookalikes. These results meet our objectives of providing accurate diagnostic and prognostic tools using noninvasive biomarkers which can be further analyzed to potentially examine pathophysiologic differences. The connectivity signatures used in these models may be further exploited to look at specific brain communication pathways associated with better or worse prognoses by our models, though we do not perform that feature analysis here. These results confirm previous suggestions of a robust functional signal of PD pathophysiology, and extend the generally-held notion of univariate correlates to longitudinal prognosis (Burciu et al., 2016; Hou et al., 2017; Manza et al., 2016; Simioni et al., 2016) to a multivariate predictive model. The confounding effect of medication can obfuscate the results of such longitudinal studies, while our results mitigate this effect by using fMRI and MDS-UPDRS scores acquired while the patients were weaned overnight from medication use.

Across all predictive targets, we found no major trend in terms of which connectivity measure provided the best predictive performance, with the exception of Partial Correlation and SP.GC potentially being less prone to overfitting, as other metrics did perform worse than chance at high augmentation levels. Partial correlation and SP.GC never dropped below chance performance, regardless of the level of augmentation and therefore may be slightly better than other measures in this case. The differences in final performance between the best Partial Correlation and best SP.GC models was not significant. The similarity in performance could indicate that all connectivity metrics tested converge on a common fMRI signal, or it could be due to another factor. For all predictive targets, we also found that after a large amount of augmentation (an augmentation factor of 10 or 20), some models test performances would break the general trend of improving with more and more augmentation and perform closer to chance. This is potentially due to failue of the Bayesian hyperparameter optimization to find a high performing area in hyperparameter space when the number of synthetic samples is much greater than the number of original samples.

Augmentation provided minimal systematic benefit to the performance of the diagnostic task, and the choice of connectivity measure also had little effect. However, the highest performance on the diagnosis was achieved with an augmentation factor of 10, suggesting that there may still be some small benefit. Our second experiment to predict the longitudinal progression of MDS-UPDRS had somewhat different results. We found augmentation to be necessary to form accurate predictors. We did not find any benefits to 'hypermatching', but did find that an augmentation factor of 3 or more tended to produce more accurate results than other approaches. For our third experiment to predict longitudinal changes in MoCA, we found very similar results as in our second experiment – that augmentation was necessary and that the choice of connectivity measure had only a small effect. However, the benefits of augmentation also waned as the total amount of augmentation increased.

### 5.2 Results in Context

Our first experiment on predicting the differential diagnosis of PD versus close lookalikes achieved a high balanced accuracy of 0.68, which is similar to current clinical accuracy, but does not exceed the best performance achieved by Zhang et al (Zhang et al., 2020) of 83% balanced accuracy, as shown in **Table 3**. This suggests that the diagnostic signal present in the 3T fMRI connectivity features is weaker than the signal present in other modalities such as diffusion weighted MRI, used by Chougar et al. or T1 MRI used by Zhang et al. Further experiments can be conducted in the future to identify if the signal in connectivity, DTI, and T1 are complimentary or fully redundant. However, we still have developed a highly accurate diagnostic model that is approximately equal in performance to standard diagnostic measures. The standard general clinical balanced accuracy is 63%, and 87% in specialist clinics (Andrew J. Hughes, Susan E. Daniel, Yoav Ben-Shlomo, Andrew J. Lees, 2002). More work needs to be done for this diagnostic task.

Work on predicting the longitudinal progression of MDS-UPDRS and MoCA has been less successful. Previous attempts to perform these longitudinal measures of prognosis have found MAEs in the longitudinal prediction of MDS-UPDRS-III of 3.22 to14.0 and $R^2$s of 0.44 to 0.56, using fMRI, SPECT, genetics, and T1 MRI. Using the recently proposed connectivity measures we achieve an MAE of 1.8, an $R^2$ of 0.69, and a p-value of 6.0E-7. The longitudinal prediction of MoCA is comparatively understudied. While studies have predicted the longitudinal progression of PD patient dementia status from cognitively normal to minor cognitive impairment (Lin et al., 2021; Silva-Batista et al., 2018), but there is a dearth of attempts to regress PD cognitive levels, such as measured by MoCA, longitudinally. Only Zeighami et al

**Table 3: Comparison to the literature.** The performance of similar diagnostic and prognostic models in the literature is compared to the performances achieved in this paper. The performance of the best Partial Correlation and best SP.GC models was not significantly different from each other, but both are significantly better than competing models. The source, kind of data, modeling approach, and final performance for each diagnostic and prognostic model are shown. This paper's results are highlighted in grey and the best performances are bolded.

| Target | Author | Data | | | | Model | Performance | | |
|---|---|---|---|---|---|---|---|---|---|
| | | *Subjects* | *Feature* | *Scan Type* | *On Meds* | | *Balanced Acc* | | |
| **Differential diagnosis** | (Chougar et al., 2021) | 228 | Regional values | DTI and T1 | NA | Logistic regression | 0.77 | | |
| | (Zhang et al., 2020) | 363 | Regional values | T1 | Yes | Univariate | ***0.83*** | | |
| | This Paper | 146 | Clinico-demo | rest fMRI* | No | XGBoost | 0.68 | | |
| | | | | | | | *MAE* | *$R^2$* | |
| **MDS-UPDRS-III prognosis** | (Nguyen et al., 2021b) | 82 | ReHO & fALFF | rest fMRI | Yes | Logistic regression | $14^{\dagger\P}$ | 0.56 | |
| | (Chahine et al., 2019) | 413 | Volumetry + CSF | T1 | Yes | LME | NA | 0.44 | |
| | (K. H. Leung et al., 2019) | 198 | Regional values | SPECT | Yes | Deep learning | 3.22 | NA | |
| | (Son et al., 2016) | 100 | Genetics + regional | SPECT | Yes | Linear regression | $8.36^{\dagger\P}$ | NA | |
| | This Paper | 63 | Partial Corr | rest fMRI* | No | XGBoost | ***1.8*** | ***0.69*** | |
| | | | SP.GC | | | | ***2.0*** | ***0.59*** | |
| | | | | | | | *MAE* | *$R^2$* | *Dx acc* |
| **MoCA prognosis** | (Lin et al., 2021) | 131 | Regional values | DTI | Yes | Random Forest | NA | NA | 83.9 |
| | (Zeighami et al., 2019) | 222 | Volumetry | T1 | No | Univariate | $0.74^{\dagger\dagger}$ | $0.06^{\S}$ | NA |
| | (Silva-Batista et al., 2018) | 39 | Balance scale | Clinic | Yes | Univariate | NA | $0.52^{\P}$ | NA |
| | This Paper | 71 | Partial Corr | rest fMRI* | No | XGBoost | ***0.60*** | ***0.81*** | NA |
| | | | SP.GC | | | | ***0.61*** | ***0.79*** | NA |

* = task-based treated as rest, † = RMSE, †† = most severe tertile versus least severe only, ¶ = current (not change over time) , § = r provided, R2 approximated

2019 (Zeighami et al., 2019) has made an attempt, and their attempt used only the least and most severe tertiles for their prediction, dropping the more difficult to predict inner tertile. We achieve a superior performance including both a lower MAE and higher $R^2$ than previous attemts. A comparison of those methods to the method presented in this paper are also shown in **Table 3**.

## 5.3 Limitations and future directions

This work uses the PDBP dataset because it mitigated the levodopa medication confound by acquiring all measures including imaging and symptomatology in the OFF medication state. PDBP is the largest database of PD with OFF state measures and has been relatively understudied. However, a larger with 5x to 10x the number of subjects would likely be more representative of the natural heterogeneity in parkinsonism,and likely allow training a higher performing models more likely to generalize to disparate patient populations. For increased clinical utility, it would be useful to extend the models developed herein to predict multiple years ahead, with data from a multi-year longitudinal study rather than a single year. Additionally, longer fMRI acquisitions of 15 min or more with true resting state, rather than task treated as rest, would likely provide superior results. It is likely that differences in performances across connectivity measures will become more apparent as the length of the fMRI acquisition increases. Finally, external validation on an additional dataset would increase the confidence in the results, so additional OFF medication subjects acquired would allow for more testing and validation of these results and models.

## **6** CONCLUSIONS

We have used a combination of classical and novel, causal measures of brain connectivity in combination with a framework for fMRI timeseries data augmentation to achieve state of the art prediction accuracy in predicting PD severity in several domains. When predicting the 1 year change in MDS-UPDRS and 1 year change in MoCA for PD patients, we achieve an MAE of 2.0 and 0.60 respectively. These results surpass thresholds of clinical utility with an MRI modality readily available in many clinical settings. We also identify specific connections with a multivariate relationship to better or worse prognosis in PD. Our results confirm the presence of early, sensitive connectivity biomarkers of the progression of Parkinson’s disease and hold great potential for furthering the pathophysiological understanding of PD and hopefully for providing patients and clinicians a much-needed prognostic indicator where none has previously existed.

## **7** ACKNOWLEDGEMENTS

Cooper Mellema was supported by NIH NINDS F31 fellowship NS115348. Alex Treacher and Albert Montillo were supported by NIH NIA R01AG059288. Albert Montillo was additionally supported by NIH NCI U01 CA207091, the King Foundation, and the Lyda Hill Foundation. Figures for this manuscript were generated with the Matplotlib and Seaborn packages (Hunter, 2007; Michael L. Waskom, 2021). Analysis was performed with the XGBoost, sklearn, and Statsmodels packages (Chen and Guestrin, 2016; Pedregosa et al., 2011; Skipper Seabold, 2010).

## **8** MATERIALS AND ETHICS STATEMENT

To facilitate reuse and extension, the authors are pleased to provide full source code for the connectivity measures and the analyses of this manuscript at the time of publication. The PDBP dataset used for this study is publically available at (https://pdbp.ninds.nih.gov/). The data used for this study were deidentified in accordance with NIH and HIPPA guidelines. All data was gathered with informed consent from all participants.

# 10 SUPPLEMENT

## 10.1 Abbreviations

A summary table of abbreviations used throughout this manuscript is described in **Supplemental Table S1**.

**Table S1**: Table of abbreviations used throughout the manuscript.

| Abbreviation | Description |
|---|---|
| **MRI** | Magnetic Resonance Imaging |
| **fMRI** | Functional Magnetic Resonance Imaging |
| **PET** | Positron emission tomography |
| **SPECT** | Single photon emission computed tomography |
| **BOLD** | Blood-oxygen Level dependant |
| **dMRI** | Diffusion Magnetic Resonance Imaging |
| **DTI** | Diffusion tensor imaging |
| **FC** | Functional Connectivity |
| **EC** | Effective Connectivity |
| **GC** | Granger Causality |
| **PD** | Parkinson's Disease |
| **PSP** | Progressive supranuclear palsy |
| **MSA** | Multiple system atrophy |
| **NC** | Normal control |
| **PDBP** | Parkinson's Disease Biomarker Project |
| **Corr** | Correlation |
| **PCorr** | Partial Correlation |
| **XGB or XGBoost** | Extreme Gradient Boosting |
| **PCA** | Principal components analysis |
| **$\text{ML.FC}_{\text{XGB}}$** | Machine Learning Functional Connectivity with an XGBoost model |
| **PC.GC** | Principal Component autoregressive model Granger Causality |
| **SP.GC** | Structurally Projected Granger Causality |
| **MDS-UPDRS** | Movement Disorder Society-sponsored revision of the Unified Parkinson's Disease Rating Scale |
| **MoCA** | Montreal Cognitive Assessment |
| **FDR** | False discovery rate |
| **MCI** | Minor cognitive impairment |

## 10.2 Functional MRI processing pipeline

The PDBP dataset was processed with an in-house pipeline with advanced motion correction (Raval et al., 2022). The structural T1-weighted MRI (sMRI) were first processed with the ConsNET tool to isolate the brain voxels (Lucena et al., 2019). Through a rigid body, affine, and then nonlinear SyN registration with ANTs, the images were spatially normalized to the MNI152 T1-weighted template (Avants et al., 2014; John Mazziota et. al., 2001). This method outperforms other candidate registration methods (Raval et al., 2022). The Functional MRI (fMRI) were frame-to-frame motion corrected with FSL's McFLIRT, and outlier frames were regressed out with Nipype's RapidArt (Jenkinson et al., 2002; Whitfield-Gabrieli and Nieto-Castanon, 2012; Woolrich et al., 2009). Outlier frames were identified if the intensity was >3 standard deviations from the mean intensity or if the head motion exceeded 1.0 mm. This threshold is suppored by published recommendations (Siegel et al., 2014). Brain extraction was performed with performed with fMRIPrep using than intersection of two segmentations, FSL'S BET2 and AFNI's 3dAutomask (Cox; Esteban et al., 2019; Woolrich et al., 2009). Spatial normalization to the MNI 152 EPI template was performed, which has been shown to better correct geometric distortions caused by EPI inhomogeneities than the alternative of T1-based normalization (Calhoun et al., 2017; Dohmatob et al., 2018). Motion-related artifacts were suppressed with ICA-AROMA (Pruim et al., 2015). Mean regional signals were extracted with the aforementioned Schaefer atlas with 100 cortical ROIs augmented with 35 subcortical regions from the AAL atlas (Mellema and Montillo, 2022; Rolls et al., 2020a; Schaefer et al., 2018). These mean regional signals were linearly detrended and scaled to a mean of 0 and unit variance.

### 10.3 Connectivity definitions

The FC measures used in this manuscript include Pearson's r and partial correlation. In addition, several newly proposed measures of functional and effective connectivity are used, including $ML.FC_{XGB}$, PC.GC, and SP.GC (Mellema and Montillo, 2022). These methods each measure a degree association between two mean regional timeseries.

Pearson's r measures the direct linear correlation between two sets of data – normalizing covariance between timeseries $X_1$ and $X_2$ by their standard deviations $\sigma_{X_1}$ and $\sigma_{X_2}$. The equation for Pearson's R is below in **Supplemental Equation S1**.

$$r = \frac{\mathrm{cov}(X_1, X_2)}{\sigma_{X_1}\sigma_{X_2}} \qquad \textbf{(S1)}$$

The calculation of partial correlation is closely related to Pearson's r. Partial correlation measures the degree of linear association between mean regional timeseries with confounding variables removed. The partial correlation between timeseries $X_1$ and $X_2$ while normalizing for n controlling variables $X_{n\notin 1,2}$ (in the case of neuroimaging, all other regional timeseries) is defined as the correlation between the residuals $\epsilon_{X_1}$ and $\epsilon_{X_2}$ for linear models predicting $X_1$ or $X_2$ from $X_{n\notin 1,2}$. This formulation is presented in **Supplemental Equation S2**, where f is a linear model predicting $X_i$ from $X_{n\notin 1,2}$.

$$\rho = \frac{\mathrm{cov}(\epsilon_1, \epsilon_2)}{\sigma_{\epsilon_1}\sigma_{\epsilon_2}}, \qquad \epsilon_i = X_i - f\left(X_{n\notin 1,2}\right) \qquad \textbf{(S2)}$$

**Algorithm S1: ML.FC algorithm**

| | |
|---|---|
| **Inputs:** $\mathbf{X}_t^i, (i \in [1,N], t \in [1,T]), \tau, \mathbf{f}$; | Input timeserieswith number of regions $N$, number of timepoints $T$, max lag $\tau$, timeseries predictor **f**; |
| **Output: E**; | Output effective connectivity matrix $E$; |
| **for** $i := 1$ *to* $N$ **do** *Full model fit* | For an initial regional timeseries **X** with $N$ regions fit a full model including region $i$; |
| $\mathbf{X}_t^i = \mathbf{f}(\mathbf{X}_{t-1}, \mathbf{X}_{t-2}, ..., \mathbf{X}_{t-\tau})$; | Fit a machine learning model **f** predicting activity **X** at region $i$ at time $t$ from times $t-1, ... t-\tau$; |
| **if** $f$ = *ERT or XGB* **then** | If an XGBoost or extremely random trees predictor; |
| $G(f)_j = \sum_{n=1}^{C} P(p(n) * 1 - p(n))$; | Importance score between $i$ and $j$ equals the Gini impurity for feature $j$ with probability of the data being routed down a split $p$, proportion of data routed to that split $P$, set of all nodes that use feature $j$ of $C$; |
| **else if** $f$ = *SVM* **then** | If a support vector predictor; |
| $G(f) = w \mid (min\|\|w\|\|^2 + C\sum(\eta_i)) : (y_i(w \cdot \theta(x_i)+b) \geq 1-\eta_i, \eta_i \geq 0)$; | Importance score is the weight given feature $j$ by classification vector $w$ given the support vector optimization with regularization terms $\eta$, $\theta$, $b$; |
| $\mathbf{E}_i = G(f)$; | The EC score between $i$ and $j$ is the feature importance for $j$ of the model $f$ predicting $i$; |
| **end** | |

| **Algorithm S2: SP.GC algorithm** | |
|---|---|
| **Inputs:** $\mathbf{X}_t^i, (i \in [1,N], t \in [1,T]), \Phi, \tau, \mathbf{f}$; | Input timeseries with number of regions $N$, number of timepoints $T$, transformation matrix $\Phi$ from prior-informed sparse PCA (Equation 2), max lag $\tau$, timeseries predictor $\mathbf{f}$; |
| **Output:** Effective connectivity matrix $\mathbf{E}$; | Output effective connectivity matrix $E$; |
| **for** $i := 1$ *to* $N$ **do** *Full model fit* | For the initial low dimensional regional timeseries $\theta$ fit a full model including region $j$ in the low dimensional projection; |
|   $\theta = \mathbf{X} \cdot \Phi$; | Project the initial regional timeseries $\mathbf{X}$ to structurally constrained subspace $\theta$ using transformation matrix $\Phi$ (Equation 2); |
|   $\mathbf{X}_t^i = \mathbf{f}(\theta_{t-1}, \theta_{t-2}, ..., \theta_{t-\tau}) \cdot \Phi^T$; | Fit $\mathbf{f}$ predicting activity $\mathbf{X}$ at time $t$ and node $i$ from $\theta$ at times $t-1, ... t-\tau$; |
|   **for** $j := 1$ *to* $N$ **do** *Reduced model fit* | Fit a reduced model without region $j$; |
|     $\theta' = \mathbf{X} \cdot \Phi_{j \setminus N}$; | Project into structurally constrained low dimensional space $\theta'$ with $\Phi_{j \setminus N}$, the transformation matrix with column $j$ removed; |
|     $\mathbf{X}_t^i = \mathbf{r}(\theta'_{t-1}, \theta'_{t-2}, ..., \theta'_{t-\tau}) \cdot \Phi_{j \setminus N}^T$; | Fit $\mathbf{r}$ predicting activity $\mathbf{X}$ at time $t$ and node $i$ from $\theta'$ at times $t-1, ... t-\tau$; |
|     $\mathbf{E}_{i,j} = log(\sigma(\mathbf{f}_{error}) / \sigma(\mathbf{r}_{error}))$; | The EC score between $i$ and $j$ equals the log of the ratio of the standard deviation of the residuals of the full and reduced model; |
|   **end** | |
| **end** | |

The calculation of $ML.FC_{XGB}$ is outlined in detail in (Mellema and Montillo, 2022). Row i of the $ML.FC_{XGB}$ functional connectivity matrix is populated by the feature importance of a machine learning model predicting regional timeseries $X_i$ from all other regional timeseries $X_{n \notin i}$. The Gini importance weighted by volume of samples routed through each node in an XGBoost decision model is used as the feature importance for the FC matrix. A detailed description of the calculation of $ML.FC_{XGB}$ is shown in Supplemental Algorithm S1

The calculation of EC measures in this manuscript are also outlined in detail in (Mellema and Montillo, 2022). Both Principal-Component projected Granger Causality ($PC.GC$) and Structurally-Projected Granger Causality ($SP.GC$) measures use a low-dimensional projection with a Granger-Causality scheme. Granger-causal measures define a directed edge by quantifying how the past history of signal B from a particular brain region informs the future activity of signal A, from another brain region. In neuroimaging, signal B is said to be Granger causal of signal A if a model to predict the future of A given all past information from all regions' signals including B is more accurate than a model that doesn't include B. The degree of causality is called the GC score (Granger, 1969). The $PC.GC$ and $SP.GC$ approaches project the timeseries into a lower dimensional representation and calculates a full and a reduced model in the low dimensional space before projecting the predicted activity back into the original space and calculating the error in full versus reduced models. The general low-dimensional GC schema is shown in Supplemental Algorithm S2. For $SP.GC$, we use a prior-constrained sparse-PCA projection whose objective function is shown in **Supplemental Equation S3**.

$$\vec{v_i}^* = \underset{v_i, ||v_i||=1, v_i^T, v_j=0, i \neq j, v_i \geq 0}{argmax} (\vec{v_i}^T (\mathbf{C} + \theta \cdot \mathbf{P}_i^T \mathbf{P}_i) \vec{v_i} - \lambda \cdot ||\vec{v_i}||_1), \quad \mathbf{P}_{i,j} = log(S_{i \cap j}) \qquad \textbf{(S3)}$$

**Table S2: Augmentation analysis.** This figure shows the hyperparameter ranges searched over in our XGBoost hyperparameter optimization. The parameter name is in the left column, distribution it was sampled from in the center column, and range in the right column.

| *Parameter* | *Distribution* | *Range* |
|---|---|---|
| Depth | Uniform | [3,9] |
| Learning rate | Loguniform | [0.1,100] |
| Booster | Choice | [dart, tree] |
| Tree | Choice | [aprx, hist, auto, exact] |
| Gamma | Loguniform | [0.01, 10] |
| Eta | Uniform | [0,1] |
| Min weight | Uniform | [0,2] |
| Max delta | Uniform | [0,2] |
| Subsample | Uniform | [0.1,1] |
| Alpha | Loguniform | [1e-6,1e2] |
| Lambda | Loguniform | [1e-6,1e2] |
| Tree node sampling | Uniform | [0.4,1] |
| Tree level sampling | Uniform | [0.4,1] |
| Number of estimators | Uniform | [50,500] |

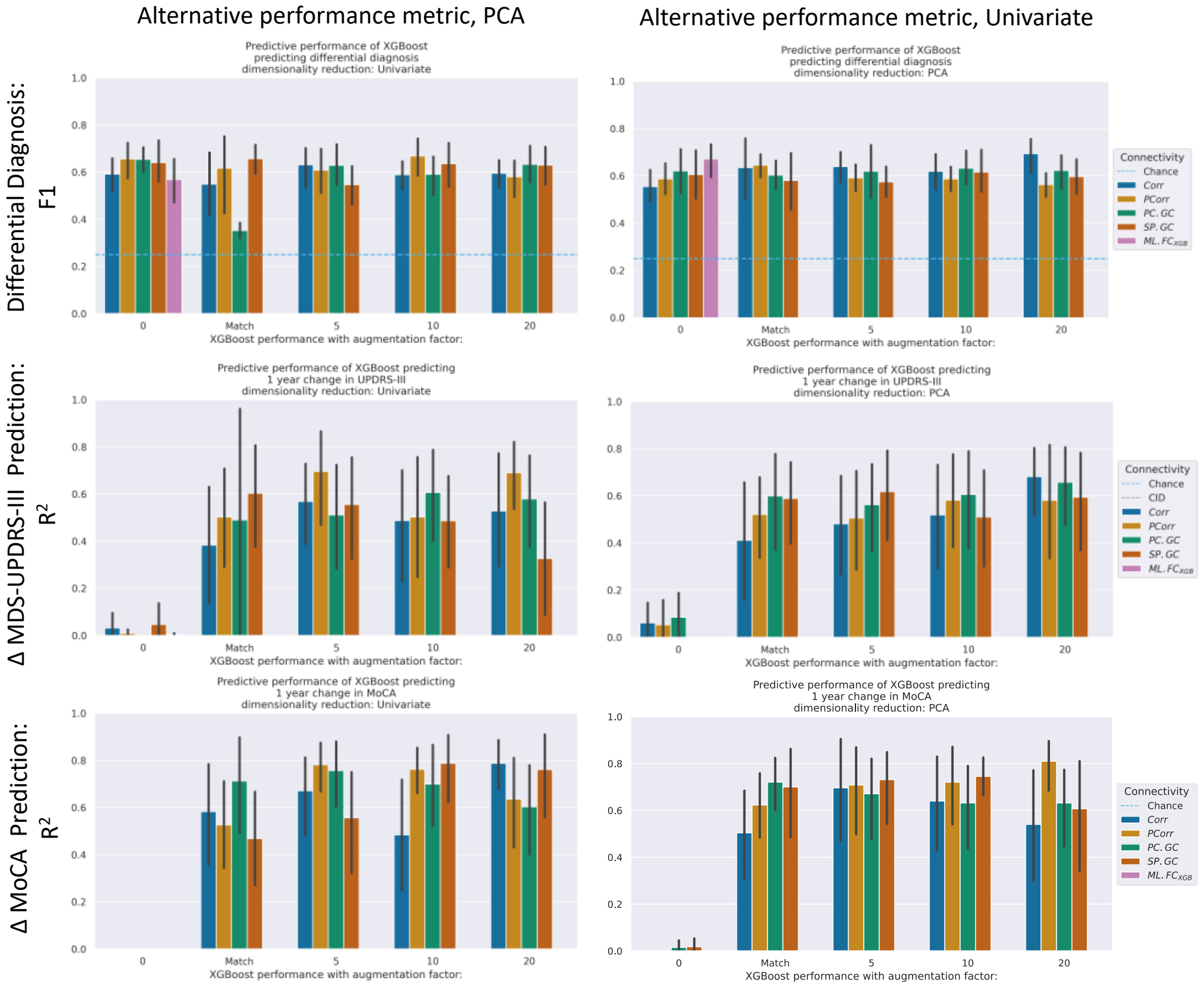

**Figure S1: Additional performance metrics.** The additional performance metrics (F1 score for classification and $R^2$ for regression) corresponding to the same models as presented in **Figure 3** are presented here. For each graph, the amount of augmentation increases along the X axis, and the performance metric is reported along the y axis. Rows are the prediction target, and columns are the dimensionality reduction technique used in the regression analysis. The 95% CI across the ten outer test folds is also shown.

### 10.4 Extreme gradient boosting model hyperparameter ranges

For Experiments 1-5, the range of hyperparameters searched remained constant for the diagnostic and prognostic predictions. The hyperparameter sampling ranges for the Bayesian optimization are shown below in **Supplemental Table S2**. The Bayesian hyperparameter optimization was performed with the python ray.tune package and python XGBoost packages.

### 10.5 Additional performance metrics for PD prediction

In addition to those presented in the results section of the paper above, additional performance metrics for the trained models were calculated. **Supplemental Figure S4** shows those additional performance metrics. For the differential diagnosis task, F1 scores are also reported. For the regression task, $R^2$ scores are also reported.

### 10.6 Augmentation sensitivity experiments

#### 10.6.1 *Experiment descriptions*

In order to examine how much augmentation aided the model in learning the prediction targets, we performed two additional subsets of experiments, one where we slowly increased the amount of augmentation from baseline and one where we enhance the amount of augmentation on the 'edge' of the score distributions analogous to curriculum learning (Han et al., 2018). For each of these models, we use the $SP.GC$ feature set with PCA dimensionality reduction in the regression. We regress both the 1 year changes in MDS-UPDRS-III and MoCA in the sensitivity analysis.

For the first augmentation sensitivity experiment, we lock the connectivity measure and dimensionality reduction technique. We chose to use $SP.GC$ and the PCA dimensionality reduction technique because it had the best performance on average across the multiple augmentation factors tested. It also was the only approach to not sometimes fail to train an accurate model (no better than chance) with high amounts of augmentation. We tested the performance of a predictive model trained with no augmentation, augmentation factors of 1, 2, 3, 5, and the 'distribution matching' augmentation factor described above. The amount of augmentation for the distribution matching scheme described in section 2.2.2 was approximately equal to an augmentation factor of 4. Each of these augmented sets was used to train a set of XGBoost models in a fashion identical to that described above.

For the second augmentation sensitivity experiment, we try an uneven matching strategy inspired by curriculum learning (Han et al., 2018) wherein we augment the edges of the distributions described in Section 2.2.2 more than the center. We attempted two versions of this 'hypermatching' and compare their performances to an evenly matched distribution. We again use the SP.GC connectivity metric with a PCA dimensionality reduction step in the regression model. For the first hypermatching scheme, we generated a "V" shaped distribution where the edges were augmented by a factor of 20 and linearly decayed to the center of the distribution. We generated a "Flat" distribution with the same number of subjects as the V-shaped distribution, but evenly distributed between bins. We also to generated a "U" shaped distribution which is in between the "Flat" and "V" distributions. Each of the Flat, U, and V shaped distributions was used to train another set of XGBoost models as described in previous experiments, using the same hyperparameter optimization scheme and nested cross-validation folds. The 'Flat', 'U', 'V', and all distributions together are displayed in **Figure S2**, in **Figure S2A, Figure S2B, Figure S2C,** and **Figure S2D** respectively.

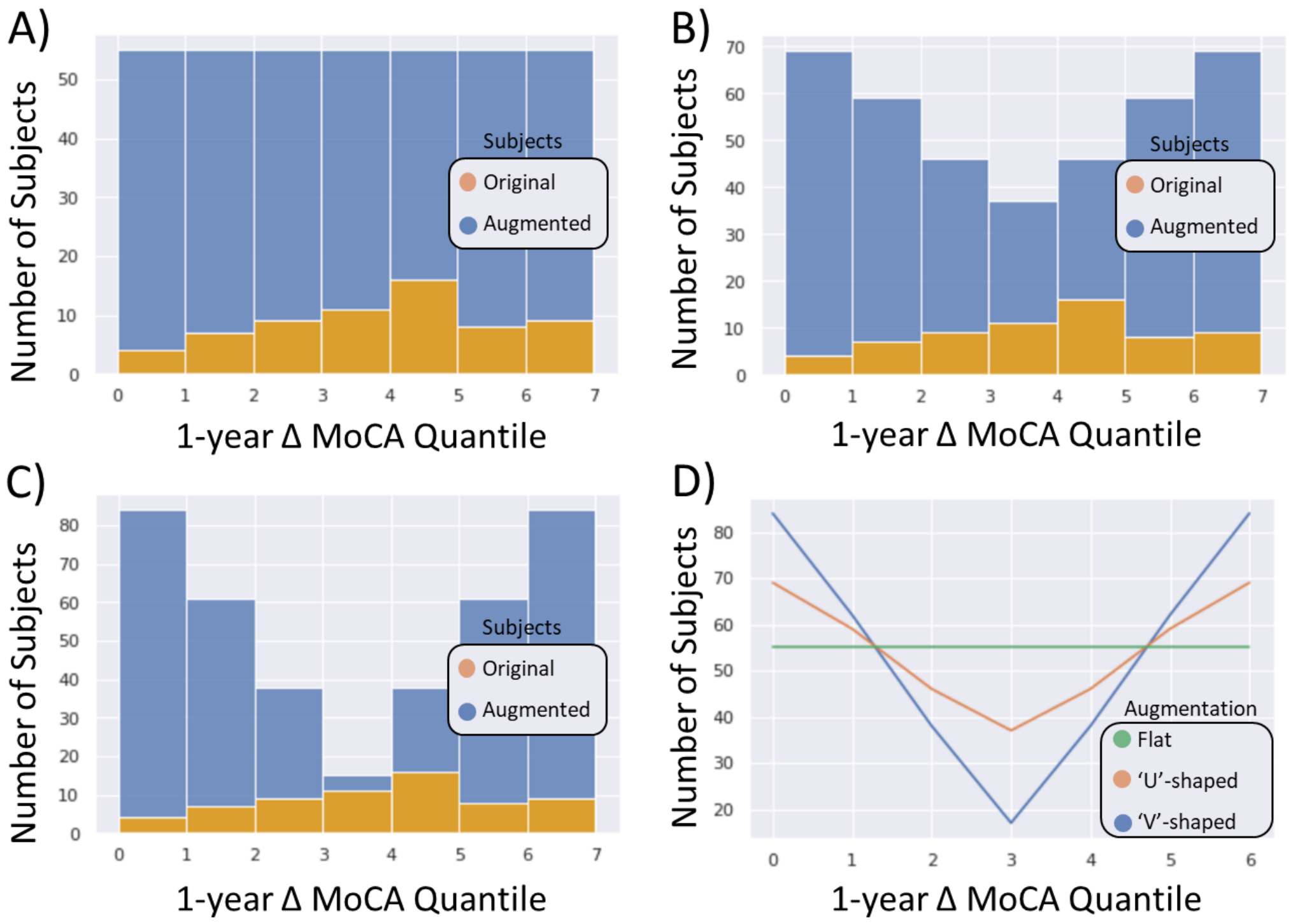

**Figure S2: Augmentation matching schematic** shows the different matching and 'hypermatching' schema inspired by curriculum learning used in the augmentation sensitivity analyses. (A) shows how the original distribution (orange) is augmented to be flat with the additional samples added to it (blue). (B) shows the 'U' shaped distribution, where the edges are amplified more than the center. (C) shows the 'V' shaped distribution, which is a more extreme version of the 'U' shaped distribution. The total number of samples is the same for all cases (A-C). The distributions plotted on top of each other are shown in (D).

#### 10.6.2 *Augmentation sensitivity experiment results*

In addition to generating high-performing models, we additionally wished to determine where the largest benefits of our augmentation approach were obtained. The performance of a predictive model using the $SP.GC$ features and PCA dimensionality reduction technique with variations to augmentation is shown in **Figure S3**. The first experiment (Experiment 4) where we vary only the augmentation amount is shown in blue, while the second experiment where we test a curriculum-learning inspired 'hypermatching' approach is shown in orange. The columns of **Figure S3** show the two predictive tasks, predicting the change in MDS-UPDRS in the first column and the change in MoCA in the second column. The first row of **Figure S3** shows the predictive performance measured with $R^2$, while the second shows the performance as measured by MAE. Note that the scales of the performance metrics are different per quadrant, and that a *higher* $R^2$ is indicative of superior performance while a *lower* MAE is indicative of superior performance. **Figure S3a** shows an increase in $R^2$ with more augmentation, with the most benefit of augmentation coming immediately with an augmentation factor of 1 (i.e. generate 1 new sample per original sample, or an increase in dataset size of 2x). **Figure S3c** shows the same pattern, but measured with MAE. **Figure S3b** and **Figure S3c** show a similar increase in $R^2$ and decrease in MAE with increasing augmented dataset size, but the largest gains in performance aren't reached until an augmentation factor of 2 or 3 (3-4x the original dataset size).

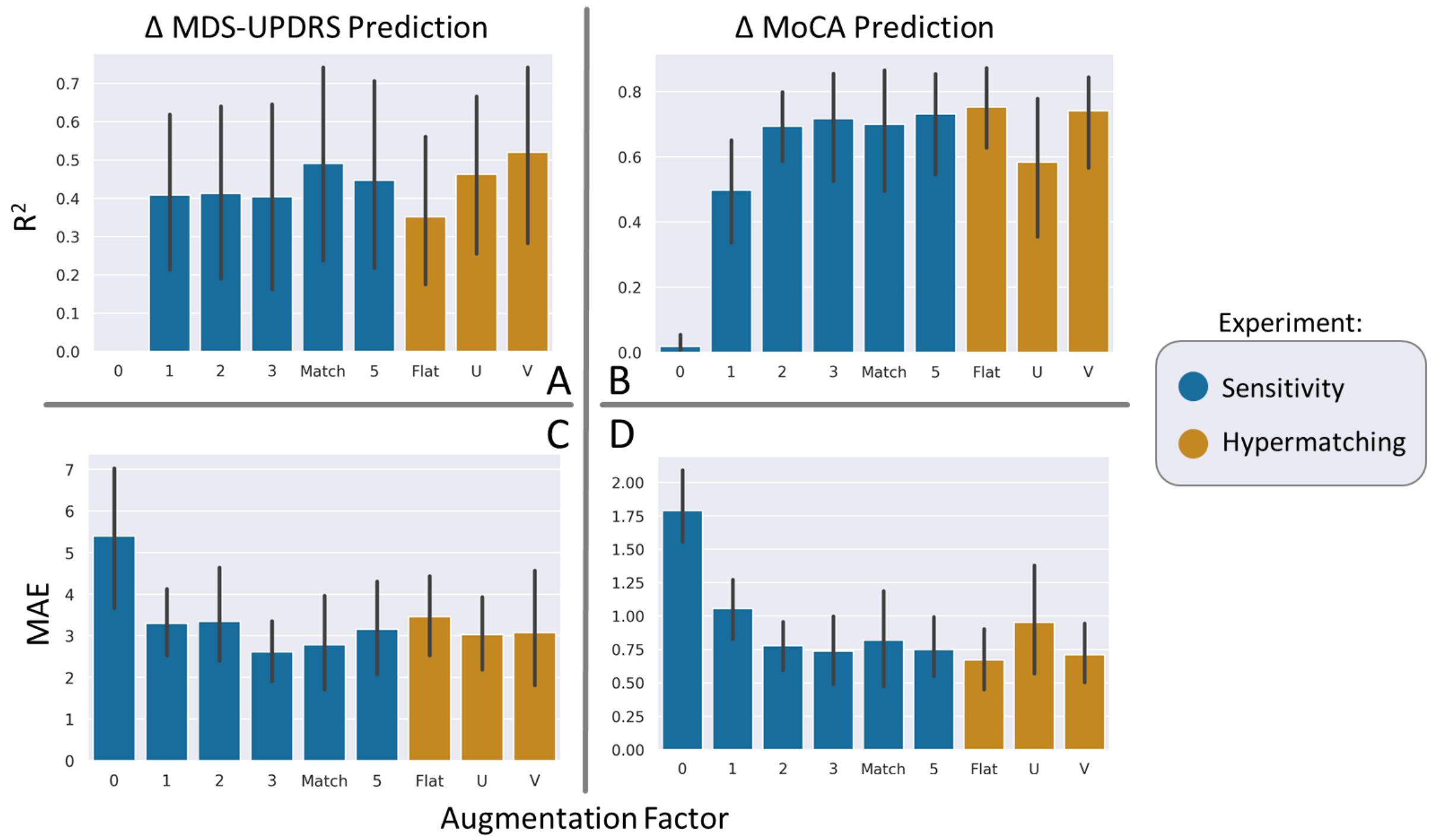

**Figure S3: Augmentation analysis.** This figure shows the results of the augmentation sensitivity analysis. Increasing augmentation is shown along each axis in blue, and the different 'hypermatching' schemes are shown in orange. (A) corresponds to the $R^2$ across the 10 outer test folds of models trained to predict the 1-year change in MDS-UPDRS with 95% CI shown. (B) corresponds to the $R^2$ across the 10 outer test folds of models trained to predict the 1-year change in MoCA with 95% CI shown. (C) corresponds to the MAE in the MDS-UPDRS predictors across those 10 outer test folds. And (D) corresponds to the MAE in the MoCA predictors across those folds.

For the second experiment with curriculum-learning inspired 'hypermatching', we find little to no difference between the various approaches. When predicting the 1 year change in UPDRS, a slight trend towards better performance with a more 'V' shaped distribution may be present, but it is not statistically significant. When predicting the 1 year change in MoCA, there is no significant difference in the different shaped augmentation patterns.

### 10.7 PCA visualization alternate views

To supplement the visualizations presented of the principal component most important for prognosis, the most important raw edges to that component are shown in **Supplemental Figure S4**. The top 2.5% of contributing edges to the component are shown, with edges with higher contribution shown in yellow and lower contribution in blue.

**Figure S4: Principal component edge visualization.** An additional visualization of the principal components most important for prognosis are presented here. This supplements the visualization presented in **Figure 4 E-F**. The edges of the connectivity graph which contribute the most to the principal components in question are shown below. The top 2.5% of edge contributions for the Partial Correlation PC and the SP.GC PC are shown. Higher contributions are shown in yellow and lower contributions are displayed in blue.

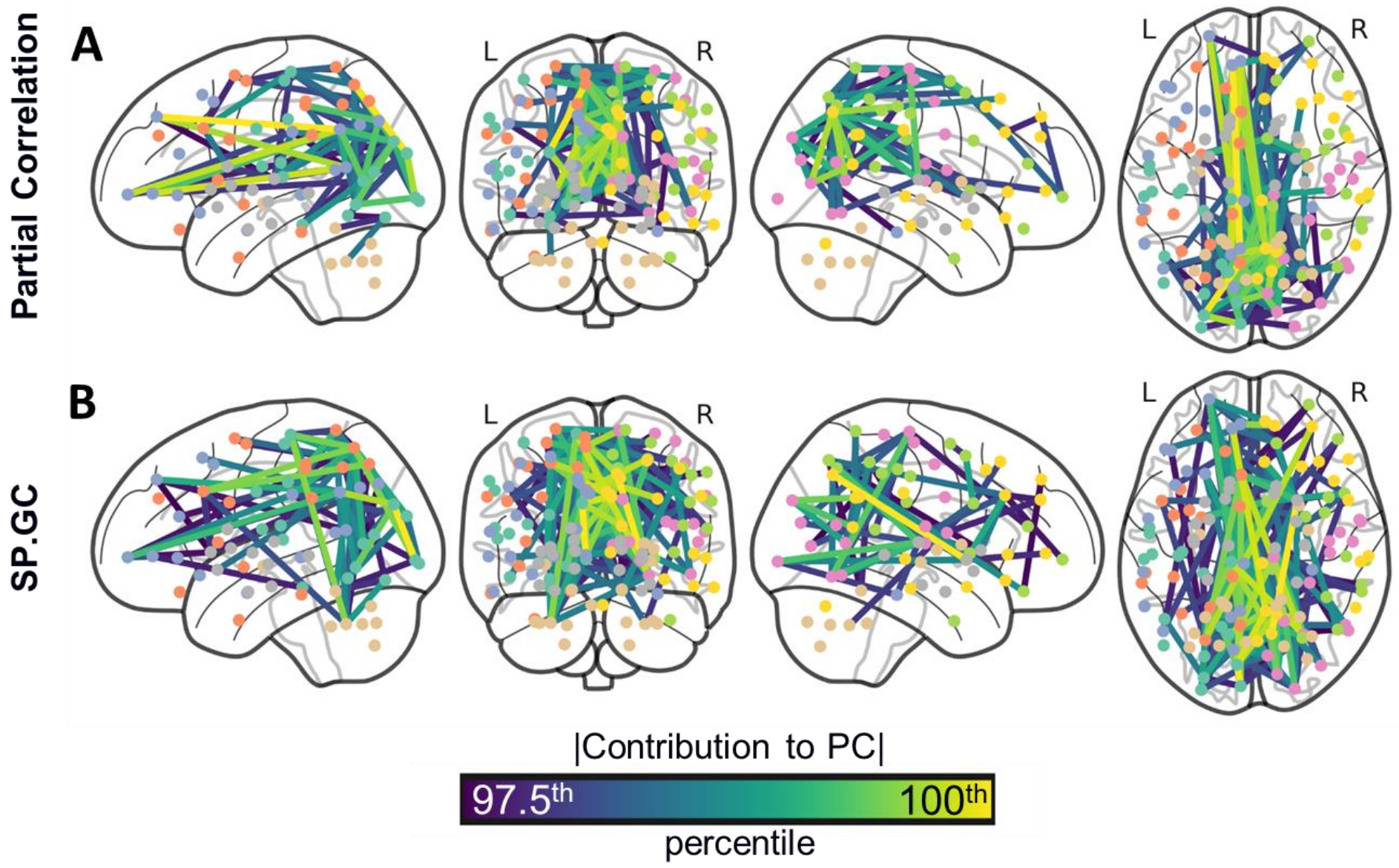